\documentclass[twocolumn,showpacs,preprintnumbers,amsmath]{revtex4}
\usepackage{graphicx}
\usepackage{amssymb}
\usepackage{bm}

\newcommand{\msun}{\mbox{$M_\odot$}}

\def\be{\begin{eqnarray}}
\def\ee{\end{eqnarray}}
\def\lsim{\mathrel{\rlap{\lower3pt\hbox{\hskip1pt$\sim$}}
     \raise1pt\hbox{$<$}}} 
\def\gsim{\mathrel{\rlap{\lower3pt\hbox{\hskip1pt$\sim$}}
     \raise1pt\hbox{$>$}}} 

\begin{document}
\setcounter{page}{1}
\title[]{Double Neutron Star Binaries: Implications for LIGO}
\author{Chang-Hwan \surname{Lee}}
\email{clee@pusan.ac.kr}
\affiliation{Department of Physics, Pusan National University, 609-735, Korea\\
Asia Pacific Center for Theoretical Physics, POSTECH, Pohang 790-784, Korea}
\author{Gerald E. \surname{Brown}}
\email{Ellen.Popenoe@sunysb.edu}
\affiliation{Department of Physics and Astronomy, State University of New York,
Stony Brook, NY 11794, USA}
\date[]{Received September 30 2005}

\begin{abstract}
Double neutrons are especially important because they give 
most accurate informations on the masses of neutron stars. 
Observations on double neutron stars show that all
masses of the neutron stars are below 1.5$\msun$.
Furthermore, two neutron stars in a given double pulsar are nearly equal in mass.
With hypercritical accretion, we found that the probability of
having companion mass $>1.5\msun$ is larger than 90\%, while
there is no observations on such systems. We believe that 
those companions with masses higher than $1.5\msun$ went into black holes,
which is consistent with our preferred maximum neutron star mass
$M_{NS}^{max} \approx 1.5\msun$ due to the kaon condensation.
In this work, we point out
that the black-hole neutron star binaries are 10 times more dominant
than double neutron star binaries. As a result, black-hole, neutron
star binaries can increase the LIGO detection rate by a factor 20.
\end{abstract}



\maketitle

\section{INTRODUCTION}




The observational evidence as well as
the calculational basis, is strong that
neutron star (NS) binaries evolve from double helium stars,
avoiding the common envelope evolution of the standard scenario \cite{vdHvP}.
(A helium star results in a giant when the hydrogen envelope is
lifted off - in binary evolution by being transferred to the 
less massive giant
in the binary.)
In the standard scenario of binary NS formation
after the more massive giant transfers its hydrogen envelope to the
companion giant, the remaining helium star burns and then explodes into a
NS. In about half the cases the binary is not
disrupted in the explosion. The NS waits until the
remaining giant evolves (and expands)
in red giant following its main sequence hydrogen burning.
Once the envelope is close enough to the NS, the
latter couples to it hydrodynamically through gravity.
Some of the material in the envelope is accreted onto the NS, although
most flies by in the wake, being heated in the process, and is lost
into space. The energy to expel the matter comes from the drop
in potential energy as the orbit of the NS tightens. 
Formulas for the
tightening and the amount of mass accreted by the NS 
were given by Bethe \& Brown \cite{BB98}.

Chevalier first estimated that in common envelope
evolution the NS would accrete sufficient matter to evolve
into a black hole (BH) \cite{Chevalier93}. 
Brown suggested the double helium star scenario,
in order to save the first born NS \cite{Brown95}. In this scenario,
mass exchange of the hydrogen envelope takes place while both
stars burn helium. There is not sufficient time for this mass 
to be accepted by either star \cite{BraunLanger}
and it is lost into space, leaving a binary of helium stars.
The two giant progenitors must have main sequence masses within
$\sim 4\%$ of each other, in order to burn helium at the
same time, a highly restrictive requirement. However, the observations
of nearly equal masses of the two NS's within the
binaries give support to this scenario, as we show.

The above scenario for binary evolution 
was made quantitative by Bethe \& Brown \cite{BB98}
who calculated that in a typical case, the NS would accrete
$\sim 1\msun$, taking it into an $\sim 2.4\msun$ 
low mass black hole (LMBH),
similar to the BH we believe resulted from SN1987A,
although somewhat more massive than the latter.
Thus, if the NS had to go through common envelope evolution
in a hydrogen envelope of $\gsim 10\msun$ from the giant, it
would accrete sufficient matter to go into a LMBH.
Therefore, when the giant evolved into a helium star, which later
exploded into a NS, a LMBH-NS binary would result provided the
system was not broken up in the explosion. (About 50\% of
the time the system survives the explosion.)
In this work, we extend Bethe \& Brown \cite{BB98} work to include
the hypercritical accretion during both red-giant and super-giant
stages of the second star which evolves later.

The above scenario was estimated
to take place 10 times more frequently than binary NS
formation which required the two stars to burn helium at the same time,
and, because of the greater mass of the BH,
the mergings of binaries with LMBH to be twice as likely to be
seen as those with only NS's. This is the origin of the factor
20 enhancement of gravitational mergers to be observed at LIGO,
over the number from binary NS's alone.


\section{Double Neutron Stars}

We list in Table~\ref{tab1} the 5 observed NS binaries with measured
masses.
Note that they are consistent with our preferred maximum mass of 
neutron star $M_{NS}^{max}=1.5\msun$ due to the kaon
condensation \cite{kaon}.
In addition, the two
NS's in a given binary have very nearly the same mass,
as would follow from the double helium star scenario,
as we discuss in next section.

\begin{table}[h]
\caption{5 observed NS binaries with measured masses.
}
\label{tab1}
\begin{tabular}{llll}
\hline
Object      & Mass ($\msun$) & Companion Mass ($\msun$) & 
Refs.\\
\hline
J1518$+$4904           & 1.56$^{+0.13}_{-0.44}$ & 1.05$^{+0.45}_{-0.11}$ & \cite{Thorsett,Nice}\\
B1534$+$12           & 1.3332$^{+0.0010}_{-0.0010}$ & 1.3452$^{+0.0010}_{-0.0010}$ & \cite{S02} \\
B1913$+$16           & 1.4408$^{+0.0003}_{-0.0003}$ & 1.3873$^{+0.0003}_{-0.0003}$ & \cite{WT03} \\
B2127$+$11C           & 1.349$^{+0.040}_{-0.040}$ & 1.363$^{+0.040}_{-0.040}$ & \cite{Deich} \\
J0737$-$3039A        & 1.337$^{+0.005}_{-0.005}$ & $^\dagger$1.250$^{+0.005}_{-0.005}$ & \cite{Lyne} \\
\hline
\end{tabular}
$^\dagger$J0737$-$3039B.
\end{table}

The very nearly equal masses of pulsar and companion in B1534$+$12
and B2127$+$11C is remarkable. We show below that B1913$+$16 comes from
a region of giant progenitors in which the masses could easily be
as different as they are. The uncertainties in J1518$+$4904 are great
enough that the masses could well be equal. The  the double pulsar
J0737$-$3039A and J0737$-$3039B were probably very nearly the same
before a common envelope evolution in which the first formed
NS J0737$-$3039A accreted matter from the evolving
(expanding) helium star progenitor of J0737$-$3039B (Case 2 in next section)
in the scenario of Dewi \& van den Heuvel \cite{Dewi}.

\section{Fate of Common Envelope Evolution}

\begin{figure}
Case 1: $\Delta T > 10 \%$, $P \sim 90\%$, $\Delta M \sim 0.9 \msun$\\
\vskip 3mm
\includegraphics[width=8.0cm]{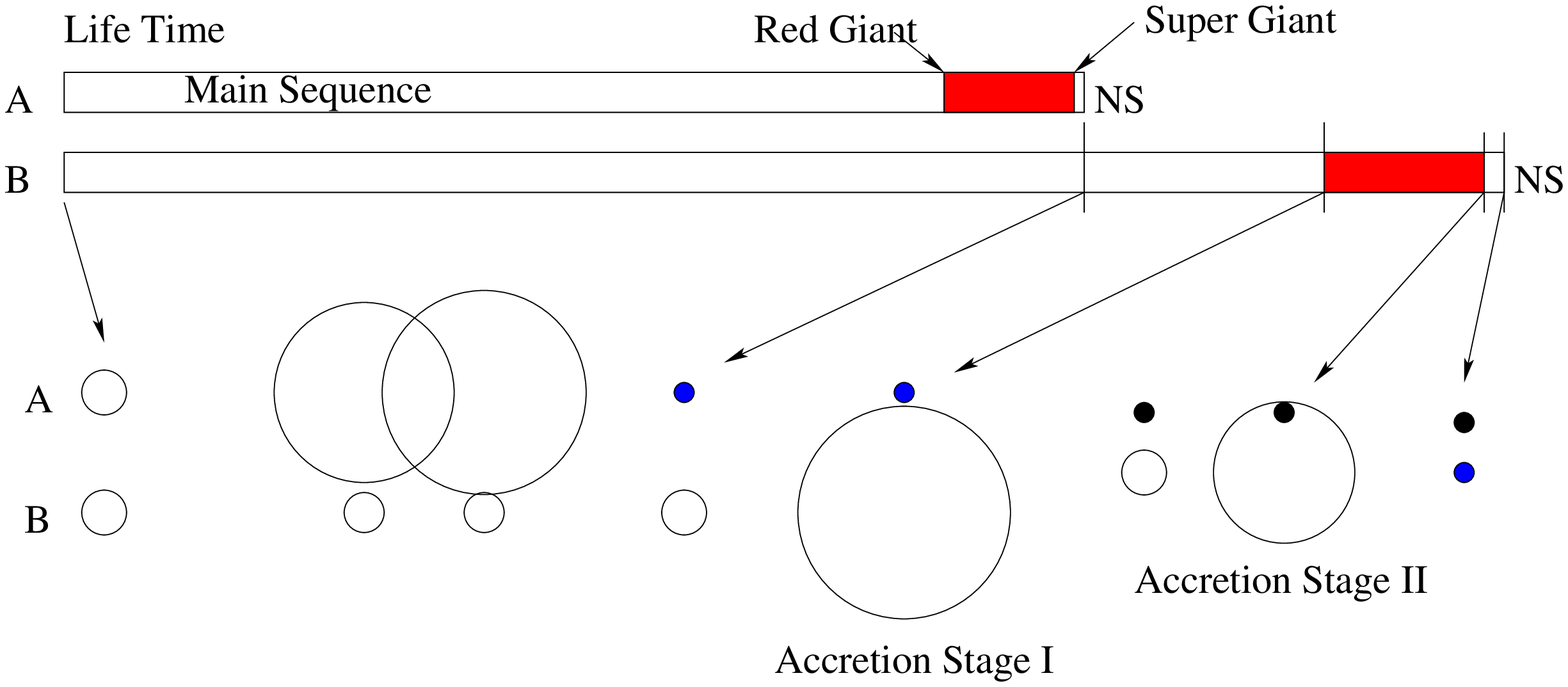}\\
\vskip 10mm
Case 2: $\Delta T < 10 \%$, $P \sim 10\%$, $\Delta M \sim 0.2 \msun$\\
\vskip 3mm
\includegraphics[width=8.0cm]{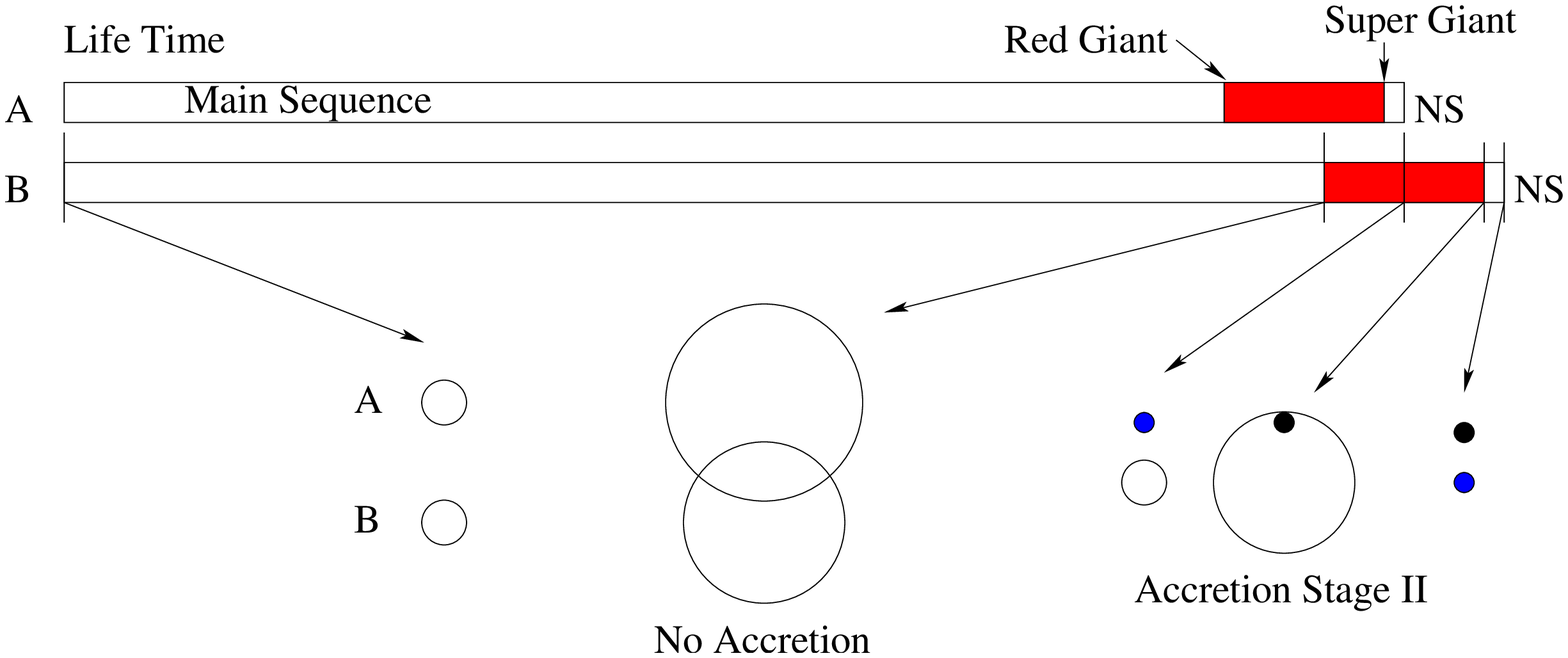}\\
\vskip 10mm
Case 3: $\Delta T < 1 \%$, $P < 1\%$, $\Delta M \sim 0.0 \msun$\\
\vskip 3mm
\includegraphics[width=8.0cm]{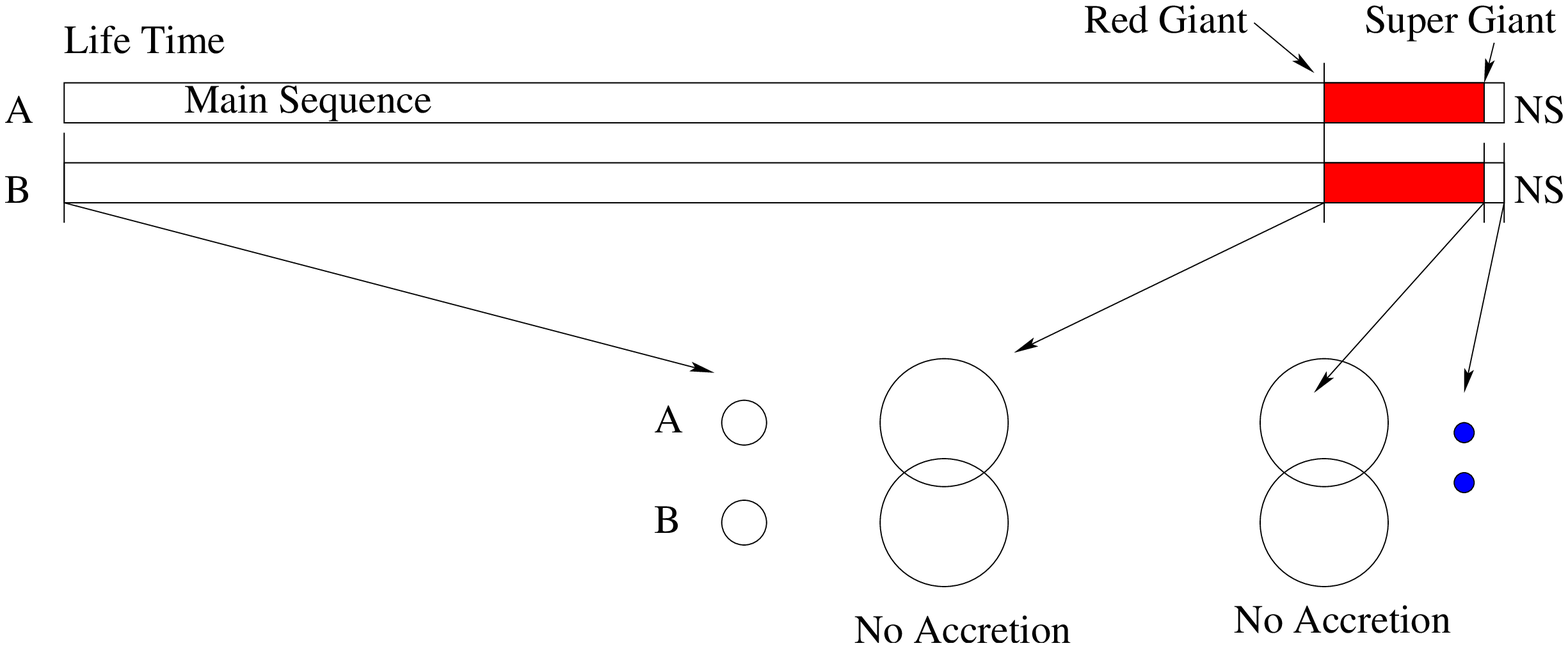}\\
\caption{Three typical cases of binary NS evolution.
$\Delta T$ is the difference in the life times of binary progenitors,
$P$ is the formation probability of binary progenitors,
$\Delta M$ is the accreted mass by the first-born NS.}
\label{fig2}
\end{figure} 

\begin{figure}[ht]
\includegraphics[width=7.0cm]{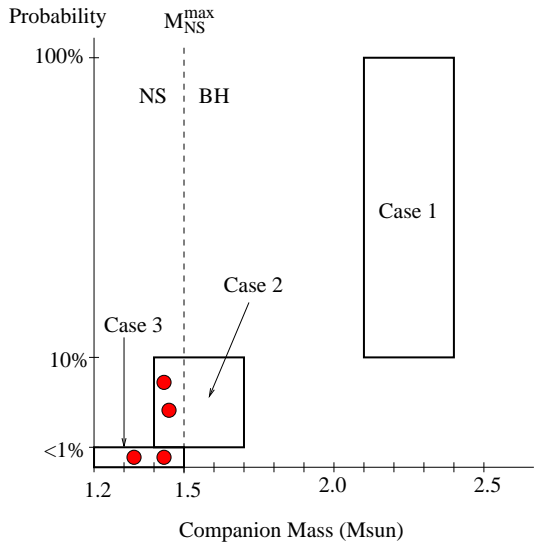}
\caption{Companion (first-born NS) mass in double NS's
and their formation probability. The area within the square
indicates the probability.
Filled circles are observed companion masses
in double NS's.
Here, NS's are assumed to be born with initial masses between
$1.2\msun$ and $1.5\msun$. 
Note that the probability of having higher mass ($>1.5\msun$) companion
is more than 90\%. We believe that they went into black holes
and
the maximum mass of NS
$M_{NS}^{max}=1.5\msun$ is consistent with observations. 
} \label{fig3}
\end{figure}

In the standard scenario the first-born NS would go through common
envelope evolution with a giant which must have ZAMS mass at least
$\sim 10\msun$ if it is to later end up as a NS \cite{vdHvP}.
In Fig.~\ref{fig2}, three typical cases of binary NS evolution
are summarized. In these estimates, we assumed that both the birth rate
and the life time are proportional to $M_{\rm ZAMS}^{-2.5}$.
With these assumption, the 4\% difference in the ZAMS mass corresponds to
10\% difference in the life time. 
Hence the population probability
for the ZAMS masses to be within 4\% difference in mass is about 10\%.

\begin{itemize}
\item
{\bf Case 1:} [90\% probability] This corresponds to the case with initial mass difference
$\Delta M_{ZAMS} > 4\%$, with life time difference $\Delta T > 10\%$.
The first born NS can accrete both
in red giant and in super giant stage of the second star which evolves later.  
Due to the hypercritical accretion,
$\Delta M = 0.9\msun$ ($0.7\msun$ in giant stage, $0.2\msun$ in supergiant stage)
can be accreted to the first-born NS. The accreted mass $\Delta M$
was estimated using the formula given in the Appendix of
Belczynski et al. \cite{Belczynski}.
\item
{\bf Case 2:} [10\% probability] This corresponds to the case with initial mass difference
$\Delta M_{ZAMS} < 4\%$, but not close enough to burn the helium at the same time.  
$\Delta M = 0.2\msun$ can be accreted to the first-born NS
during the supergiant stage of the second star.
\item
{\bf Case 3:} [$<$ 1\% probability] This corresponds to the case in which the
initial mass differences are so close to burn the helium at the same time.  
Nothing can be accreted because two NS's are formed almost at
the same time.
\end{itemize}

As in Fig.~\ref{fig3}, if the NS's are to be born with initial
masses between $1.2\msun$ and $1.5\msun$ as seen in double NS's,
the first-born NS's in Case 1 result in $2.1-2.4\msun$ due to the hypercritical
accretion. Similarly, those in Case 2 result in $1.4-1.7\msun$.
Furthermore, one shouldn't forget that these
massive pulsars in Case 1 would be copious, because there is no special
condition of the giant progenitors of the binary NS 
having to be within 4\% of each other in ZAMS mass. In fact, the
binaries with pulsar mass $2.1-2.4\msun$ would be a factor of
10 more frequent than those with companion masses $1.2-1.7\msun$. Since we
don't see pulsars with such high masses, we believe that they
must have gone into LMBH's.

Our argument in this note alone does not exclude NS in the mass range
up to $2.1\msun$, but as mentioned earlier the Bethe \&
Brown \cite{BB95} argument that the maximum mass of the
NS in 1987A, which we believe went into a LMBH,
of $1.57\msun$ further constrain the maximum NS mass,
if our belief is correct.
In Fig.~\ref{fig3}, we draw the expected maximum mass of NS
$M_{NS}^{max}=1.5\msun$ which was estimated from kaon condensation \cite{kaon}. 
Note that the probability of having higher mass ($>1.5\msun$) companion
is more than 90\% while there is no observations on such systems.
We believe that this indicates that those NS's with masses $>1.5\msun$ 
went into black holes after common envelope evolution.

The NS mass in the helium white-dwarf, NS binary
J0751$-$1807 is quoted as $2.1\pm 0.20 \msun$ \cite{nice2005}.
The NS mass in J0751$-$1807 is measured from the
period change due to gravitational wave emission. The companion white
dwarf mass is constrained by a marginal detection of Shapiro delay.
Although the observational indication of high NS mass is
strong, this mass would be brought down with the 4/3 power of
white dwarf mass if the latter were increased, and still fit the
same period change. Thus, we believe the case for such a massive
NS to only be settled with a sufficiently accurate
measurement of the Shapiro decay which pins down the white dwarf mass.
It should be noted that just in the evolution of NS,
white-dwarf binaries there is ample possibility for substantial
accretion from the evolving progenitor of the white-dwarf, so
these binaries are the place to look if one wants to find a
high-mass NS.

\section{Obervability Premium for LMBH-NS Binaries}

Why haven't we seen any LMBH-NS binaries? Van den Heuvel
\cite{vdH,vdH2}
has pointed out that NS's form with strong magnetic fields
$10^{12}$ to $5\times 10^{12}$ gauss, and spin down in a time
\be
\tau_{\rm sd} \sim 5\times 10^6 \; {\rm years}
\ee
and then disappear into the graveyard of NS's.
The relativistic binary B1913$+$16 has
a weaker field $B\simeq 2.5\times 10^{10}$ gauss and therefore
emits less energy in magnetic dipole radiation. Van den Heuvel estimates
its spin-down time as $10^8$ yrs. There is thus a premium in observational
time for lower magnetic fields, in that the pulsars can be seen for
longer times.
Taam \& van den Heuvel \cite{Taam86} found empirically that the
magnetic field of a pulsar dropped roughly linearly with accreted
mass. This accretion can take place from the companion in any stage
of the evolution.
A pulsar that has undergone accretion is said to have
been ``recycled".
In B1913$+$16 the pulsar magnetic field is
$\sim 2.5\times 10^{10}$ gauss and in B1534$+$12 it is 
$\sim 10^{10}$ gauss. In J0737$-$3039A it is only
$6.3\times 10^9$ gauss. These ``recycled pulsars" will
be observable for $\gsim 100$ times longer than a ``fresh" (unrecycled)
pulsar. 

The same holds for LMBH-NS binaries. The NS is certainly not recycled,
so there is an about 1\% chance of seeing one as the recycled pulsar
in a binary NS. But we propose 10 times more of the LMBH's than binary
NS's, of which we observe 5. Thus the total probability of seeing
the LMBH binary should be about 50\%. 
However, there may be additional reasons that the
LMBH-NS binary is not observed.

\begin{table}
\caption{Predicted LIGO Detection Rates (yr$^{-1}$).
} 
\label{tab2}
\vskip 2mm
\begin{center}
\begin{tabular}{cccc}
\hline
Binary Type & LIGO I & LIGO II  & Chirp Masses ($\msun$) \\
\hline
NS-NS$^\dagger$          & 0.0348  &  187   & 1.0 - 1.3 \\
BH-NS$^{\dagger\dagger}$ & 0.696   & 3740   & 1.3 - 2.7 \\
BH-BH$^{\star\star}$     & 0.58    & 2450   & $\sim 6$ \\
Total                    & 1.31    & 6377   & \\
\hline
\end{tabular}
\end{center}
\vskip 2mm
{\small
$^\dagger$ NS-NS detection rates are from Kalogera et al. \cite{Kalogera}.
$^{\dagger\dagger}$ BH-NS detection rates are obtained by multiplying
factor 20 to NS-NS detection rates from Bethe \& Brown \cite{BB98}.
$^{\star\star}$ BH-BH detection rates are from Portegies Zwart \&
McMillan \cite{simon} with the modification of BH mass $7\msun$.
}
\end{table}

In Table~\ref{tab2}, the estimated LIGO detection rates are summarized.
For completeness we added the contribution from BH-BH mergers obtained by
Portegies Zwart \& McMillan \cite{simon}.
They have suggested a large number of gravitational mergings of high-mass
BH's ejected from globular clusters. Their predictions should be
tested relatively early in the LIGO development.

\section{Conclusion}

The discovery of the double pulsar increased estimated rate for
gravitational merging by a factor of 6$-$7 over that of Kim
et al. \cite{Kim}. In addition to these effect,
we find 10 times more LMBH-NS binary mergings, with larger chirp
mass than NS-NS binaries because of the accretion in forming
the BH so that these mergings multiply the binary NS
ones by a factor of 20. 
According to our estimates, LIGO I would be able to detect one
mergings per year.


The Chirp mass, which will be detected with an estimated accuracy
of $\sim 0.002\msun$
\be
M_{\rm chirp}=\mu^{3/5} M^{2/5}
=\frac{(M_1 M_2)^{3/5}}{(M_1+M_2)^{1/5}}
\ee
would be $1.22 \msun$ for the merging of two $1.4\msun$ NS's,
for the merging of the least massive $2.1\msun$ BH, resulting from
the common envelope evolution of a $1.4\msun$ NS in a ZAMS $10\msun$
progenitor companion the chirp mass would be $1.49\msun$.
Estimated Chirp masses are summarized in Table~\ref{tab2}.
We may have to wait for
LIGO which will be able to measure ``chirp" masses quite accurately.
The chirp mass of a NS binary should concentrate near $1.2\msun$,
whereas the LMBH-NS systems should have a chirp mass
of $> 1.4\msun$, and there should be $\sim 20$ times more of the
latter.


\begin{acknowledgments}
CHL was supported by 
grant No. R01-2005-000-10334-0(2005) from the Basic Research 
Program of the Korea Science \& Engineering Foundation.
GEB was supported in part by the US Department of Energy 
under Grant No.DE-FG02-88ER40388.
\end{acknowledgments}

\def\ApJ{Astrophysical J.}



\begin{thebibliography}{}

\bibitem{vdHvP}
E.P.J. van den Heuvel and J. van Paradijs,
Scientific American, November, 38 (1993).

\bibitem{BB98}
H.A. Bethe and G.E. Brown, \ApJ, 506, 780 (1998).

\bibitem{Chevalier93}
R.A. Chevalier, \ApJ, 411, L33 (1993).

\bibitem{Brown95}
G.E. Brown, \ApJ, 440, 270 (1995).

\bibitem{BraunLanger}
H. Braun and N. Langer, Astronomy \& Astrophysics, 297, 483 (1995).

\bibitem{Thorsett}
S.E. Thorsett and D. Chakrabarty, \ApJ, 512, 288 (1999).

\bibitem{Nice}
D.J. Nice, R.W. Sayer, and J.H. Taylor, \ApJ, 466, L87 (1996).

\bibitem{S02}
I.H. Stairs, S.E. Thorsett, J.H. Taylor, and A. Wolszczan,
\ApJ, 581, 501 (2002).

\bibitem{WT03}
J.M. Weisberg and J.H. Taylor,
in {\it Radio Pulsars},
eds M. Bailes, D.J Nice and S. Thorsett,
93-98 (Astronomical Society of the Pacific, San Francisco, 2003)

\bibitem{Deich}
W.T.S. Deich and S.R. Kulkarni,
in {\it Compact Stars in Binaries: IAU Symposium 165},
eds J. van Paradijs, E.P.J. van den Heuvel and E. Kuulkers,
279-285 (Kluwer, Dordrecht, 1996).

\bibitem{Lyne}
A.G. Lyne, et al., Science, 303, 1153 (2004).

\bibitem{kaon}
G.Q. Li, C.-H. Lee, G.E. Brown, Phys. Rev. Lett. 79, 5214 (1997).

\bibitem{Dewi}
J.D.M. Dewi and E.P.J. van den Heuvel, Mon. Not. Roy. Astron. Society, 349, 169 (2004).

\bibitem{Belczynski}
K. Belczynski, V. Kalogera, and T. Bulik,
\ApJ, 572, 407.

\bibitem{nice2005}
D.J. Nice, et al.,
{\it A 2.1 Solar Mass Pulsar Measured by Relativistic Orbital Decay"},
astro-ph/0508050. 

\bibitem{BB95}
H.A. Bethe and G.E. Brown, \ApJ, 445, L129 (1995).

\bibitem{vdH}
E.P.J. Van den Heuvel,
Interacting Binaries: Topics In Close Binary Evolution.
in {\it Lecture notes of the 22nd Advanced Course of the Swiss Society for 
Astronomy and Astrophysics (SSAA)},
eds  H. Nussbaumer and A.  Orr, 263-474 (Berlin-Springer, 1994).

\bibitem{vdH2}
D. Bhattacharya and E.P.J. van den Heuvel, Phys. Rept., 203, 1 (1991).

\bibitem{Taam86}
R.E. Taam and E.P.J. van den Heuvel, \ApJ, 565, 235 (1986).

\bibitem{simon}
S.F. Portegies Zwart and S.L.W. McMillan, \ApJ, 528, L17 (2000).

\bibitem{Kalogera}
V. Kalogera, et al., \ApJ, 614, L137 (2004).

\bibitem{Kim}
C. Kim, V. Kalogera, and D.R. Lorimer, \ApJ, 584, 985 (2003).

\end{thebibliography}
\end{document}